\documentclass[fleqn,usenatbib]{mnras}

\usepackage{newtxtext,newtxmath}
\usepackage{ulem}
\usepackage[T1]{fontenc}

\DeclareRobustCommand{\VAN}[3]{#2}
\let\VANthebibliography\thebibliography
\def\thebibliography{\DeclareRobustCommand{\VAN}[3]{##3}\VANthebibliography}

\usepackage{graphicx}	% Including figure files
\usepackage{amsmath}	% Advanced maths commands
\usepackage[dvipsnames]{xcolor}	% Advanced maths commands

\title[Eccentricity Evolution in Gaseous Dynamical Friction]{Eccentricity Evolution in Gaseous Dynamical Friction}

\author[Ákos Szölgyén]{
Ákos Szölgyén$^{1,2}$\thanks{E-mail: akos.szolgyen@ttk.elte.hu},
Morgan MacLeod$^1$,
Abraham Loeb$^1$
\\
$^{1}$ Harvard-Smithsonian Center for Astrophysics, 60 Garden Street, Cambridge, MA, 02138, USA\\
$^{2}$ Institute of Physics, Eötvös University, Pázmány P. s.  1/A, Budapest, 1117, Hungary
}

\date{Accepted XXX. Received YYY; in original form ZZZ}

\pubyear{2022}

\begin{document}
\label{firstpage}
\pagerange{\pageref{firstpage}--\pageref{lastpage}}
\maketitle

% Abstract of the paper
\begin{abstract}
We analyse how drag forces modify the orbits of objects moving through extended gaseous distributions. We consider how hydrodynamic (surface area) drag forces and dynamical friction (gravitational) drag forces drive the evolution of orbital eccentricity. While hydrodynamic drag forces cause eccentric orbits to become more circular, dynamical friction drag can cause orbits to become more eccentric. We develop a semi-analytic model that accurately predicts these changes by comparing the total work and torque applied to the orbit at periapse and apoapse. We use a toy model of a radial power-law density profile, $\rho \propto r^{-\gamma}$, to determine that there is a critical $\gamma = 3$ power index which separates the eccentricity evolution in dynamical friction: orbits become more eccentric for $\gamma < 3$ and circularize for $\gamma > 3$. We apply these findings to the infall of a Jupiter-like planet into the envelope of its host star. The hydrostatic envelopes of stars are defined by steep density gradients near the limb and shallower gradients in the interior. Under the influence of gaseous dynamical friction, an infalling object's orbit will first decrease in eccentricity, then increase. The critical separation that delineates these regimes is predicted by the local density slope and is linearly dependent on polytropic index. More broadly, our findings indicate that binary systems may routinely emerge from common envelope phases with non-zero eccentricities that were excited by the dynamical friction forces that drove their orbital tightening. 
\end{abstract}

% Select between one and six entries from the list of approved keywords.
% Don't make up new ones.

%%%%%%%%%%%%%%%%%%%%%%%%%%%%%%%%%%%%%%%%%%%%%%%%%%

%%%%%%%%%%%%%%%%% BODY OF PAPER %%%%%%%%%%%%%%%%%%

\section{Introduction}
\label{sec:introduction} 

Gravitationally bound objects follow Keplerian trajectories in the classical two-body problem. In an ambient gaseous medium, however, orbits evolve under the friction that the gas exerts on the embedded bodies. Gaseous friction changes the orbital parameters through apsidal precession, orbital migration, eccentricity change and inclination damping \citep{Grishin2015}.
We will discuss mechanisms of gas--object interaction focusing on gaseous dynamical friction, also investigating hydrodynamic drag but neglecting gaseous feedback \citep[like e.g.][]{2016A&A...589A..10T}. Both the hydrodynamic drag force and the dynamical friction force oppose the direction of motion of the orbiting body, transferring energy and angular momentum to the gas. As a consequence, the semi-major axis shrinks and the orbital eccentricity may change as well. The resulting orbital evolution, therefore, depends on the properties of the gaseous and the parameters of the system orbit. 

The response of gaseous surroundings to the passage of a gravitating object is described by the Bondi-Hoyle-Lyttleton accretion theory \citep{1939PCPS...35..405H,1944MNRAS.104..273B,1952MNRAS.112..195B,2004NewAR..48..843E}. Further, the overdense wake left by a passing object implies a momentum exchange \citep{1971MNRAS.154..141H,1985MNRAS.217..367S,1994ApJ...427..351R,1994A&AS..106..505R,1995A&AS..113..133R,1996A&A...311..817R,1999A&A...346..861R,2016A&A...589A..10T} much like that in collisionless dynamical friction \citep{Chandrasekhar1943}. 
\cite{Ostriker1999} derived gaseous dynamical friction force formulae using time-dependent linear perturbation theory. 

More broadly, the theory of gaseous dynamical friction has applications throughout astrophysics, where gas and gravitating bodies frequently intermingle. Some particularly rich applications have included the dynamics of young stars, and clusters embedded in molecular clouds \citep[e.g.][]{2010MNRAS.402.1758S,2014MNRAS.441..919L,2014ApJ...794..167S,Antoni2019,2022arXiv220301330R};
%2006ApJ...638..369K,
%2008ApJ...679L..33K,
%2011ApJ...735...25N,
%2019ApJ...876..142K,
%2019MNRAS.490.5196C,
the interactions and capture of stars and compact objects by discs in active galactic nuclei \citep[e.g.~][]{1991MNRAS.250..505S,1993ApJ...409..592A,Narayan2000,2005ApJ...619...30M,2011ApJ...726...28B,2016MNRAS.460..240K,2017ApJ...835..165B,2017MNRAS.464..946S,2019ApJ...878...85S,2020ApJ...898...25T,MacLeod2020,2021arXiv211003741M,2022arXiv220306187J}; and
%2018MNRAS.476.4224P
%2020ApJ...903..133S,
common envelope phases \citep{1976IAUS...73...75P} in which a star engulfs its companion star, planets or a compact object within its gaseous envelope and dynamical friction drives orbital inspiral \citep[e.g.][]{Iben1993,Ivanova2013,2007ApJ...661.1192V,2015ApJ...803...41M,2016MNRAS.458..832S,MacLeod2017,2018ApJ...853L...1M,2020MNRAS.493.4861G,2020ApJ...897..130D,2020ApJ...899...77E}. Hydrodynamic drag and dynamical friction contribute to the capture and migration of planetesimals in molecular clouds
\citep[e.g.][]{2021ApJ...921..168P,2022ApJ...924...96M}; protoplanetary discs \citep[e.g.][]{Grishin2015,2016ApJ...820..106G,2019MNRAS.487.3324G} and debris around white dwarfs can also capture dust and planetesimals triggering white dwarf pollution \citep[e.g.][]{2020MNRAS.498.4005O,2021MNRAS.501.3806M}. 

Following \cite{Ostriker1999}, subsequent authors examined various aspects of orbital evolution both numerically and analytically in different astrophysical contexts \citep[see][for recent discussion of this literature]{MacLeod2017,Antoni2019}. Of particular relevance to our subsequent discussion is work that considered the effect of dynamical friction on orbits. For example, \cite{Salcedo2001} used numerical simulations to study the orbital evolution of objects in different models of gaseous spheres. Later, \cite{Kim2007} studied the gravitational torque exerted on a single perturber on circular orbits by dynamical friction of a uniform gaseous medium using a semi-analytic approach. \cite{2008ApJ...679L..33K} extended the method of \cite{Kim2007} to double perturbers on circular orbits. \cite{2014ApJ...794..167S} also studied the morphology of the binary case and the torque on the centre of mass. \cite{2016ApJ...820..106G} applied the problem of dynamical friction to intermediate size binary planetesimals, analogous to binaries in active galactic nuclei discs. \cite{Salcedo2019} examined the evolution of eccentric orbits, comparing analytic and numerical results. \cite{2019MNRAS.489.5424V} studied dynamical friction in slab-like geometries such as accretion discs and provided refined analytic expressions for force in both the supersonic and subsonic regime. \cite{Bonetti2020} considered the implications of prograde and retrograde motion in rotating media. Furthermore, \cite{2021MNRAS.507.2659G} investigated the common envelope evolution of initially eccentric binaries using hydro-dynamical simulations. They found that the eccentric orbits only partially circularize during the common envelop inspiral. Recently, \cite{2022ApJ...928...64D} have used Li\'enard-Wiechert potentials to derive force expressions for a circularly moving point mass in a gaseous medium. They found that the steady state is reached after only one sound-crossing time. \cite{2022arXiv220311227Y}, however, used hydrodynamic simulations to integrate orbital evolution in the context of planetary engulfment considering both the ram pressure and gravitational drag. Among the shared questions are the characteristic length scale over which gravitational wakes extend (defining the Coulomb logarithm term) and how accelerated, rather than linear, motion affects the forces. 

In this paper, we focus particularly on the eccentricity evolution of two-body systems. We adopt a model in which a less massive secondary object orbits in the gaseous envelope of a more massive, extended primary object. We use both (i) a simple numerical integrator to solve the equations of motion, reconstruct the orbital path, and measure the eccentricity evolution; and (ii) a semi-analytic approach to directly estimate the eccentricity evolution from the orbital parameters and drag forces. In section \ref{sec:2}, we introduce the equations of motion and our numerical technique. Then we outline our semi-analytic formalism in section \ref{sec:3}. We demonstrate the application of our formalism in a toy model first, in section \ref{sec:isothermal}, where the primary object is an isothermal sphere with a power-law density profile. Then we use both numerical and semi-analytic techniques to calculate the eccentricity evolution of a Jupiter-like planet within the stellar envelope of a red giant star in section \ref{sec:4}. We examine a series of different polytropic models and show how the eccentricity evolution depends on the polytropic density profile of the envelope. We summarize our conclusions in section \ref{sec:conclusion}.

%%%%%%%%%%%%%%%%%%%%%%%%%%%%%%%%%%%%%%%%%%%%%%%%%%

\section{Method}
\label{sec:2}

\subsection{Equations of Motion}
We start with a simple model of a two-body system where a more massive object (primary) is represented by an extended, static, spherically symmetric gaseous halo; and a less massive companion (secondary) is orbiting within the gas. In this picture, the acceleration of the secondary originates from both the gravitational force of the enclosed mass ($m_\star$) and the external force ($\vec{F}_\mathrm{ext}$) that the gas exerts on it 
\begin{equation}
\label{EoM}
    \vec{a} = -G\dfrac{m_\star}{r^2} \dfrac{\vec{r}}{r} + \dfrac{\vec{F}_\mathrm{ext}}{m} \, .
\end{equation}
Here, $G$ is the gravitational constant, $\vec{r}$ is the position and $m$ is the mass of the secondary. We only examine the orbital evolution of the secondary and neglect its back-reaction on the gas. The external forces are considered to be either the hydrodynamic drag force or the gas dynamical friction. The hydrodynamic drag force \citep{Villaver2009} is
\begin{equation}
\label{hyd}
    \vec{F}_\mathrm{hyd} = - \dfrac{1}{2}C_d R^2 \pi \rho v^2 \dfrac{\vec{v}}{v} %\vec{v}^2
\end{equation}
where $R$ is the radius, $\vec{v}$ is the velocity of the secondary and we use $C_d = 1.0$\footnote{We note that $C_d$ depends on Reynolds and Mach numbers \citep{2011ApJ...733...56P} but in the ram pressure regime it can be well approximated by a constant \citep[][]{Grishin2015}, e.g. unity for large Reynolds numbers.}. $\rho$ is the local density in the gas. The gas dynamical friction force \cite{Ostriker1999} is
\begin{equation}
\label{df}
    \vec{F}_\mathrm{df} = - 4\pi(G m)^2  \dfrac{\rho I}{v^2} \dfrac{\vec{v}}{v}
\end{equation}
where $I$ is the Mach-number dependent parameter i.e.~$I_\mathrm{sub} = 0.5 \ln \left[ (1+v/c_s)/(1-v/c_s) \right] - v/c_s$ in the subsonic regime and $I_\mathrm{sup} = 0.5\ln \left( 1 -(v/c_s)^{-2}\right) - \ln \left( r / R\right)$ in the supersonic regime. $c_s$ is the local sound speed in the gas, $r$ is the characteristic size of the system (here we adopt the separation of the secondary object).

The compactness of the secondary object determines the  relative importance of the above drag forces. \cite{Grishin2015}, for example, calculated the critical size of the secondary (given its density) below which the dynamical friction becomes the dominant external force in forming the orbital evolution. Similarly by comparing Eq.~\eqref{hyd} and Eq.~\eqref{df}, we define an approximate critical compactness (a mass to radius ratio) above which the dynamical friction is dominant over hydrodynamic drag:  
\begin{equation}
    \frac{m}{R} > \frac{v^2}{2 G} \sqrt{\frac{C_d}{2 I}} 
\end{equation}

One implicit assumption in Eq.~\eqref{hyd} and Eq.~\eqref{df} is that force depends on the local conditions (those in the vicinity of the orbiting object). In principle, this might not be satisfied if there are changes in quantities like the density over the characteristic length scale of the Coulomb logarithm \citep[see][for a more detailed description of the applicability and limits of this local formalism as compared to hydrodynamic simulations]{Salcedo2019}. 

Although we neglect including feedback as an additional source of external force in our model, it is important to note that there are certain astrophysical scenarios in which gas accretion onto the secondary potentially leads non-negligible feedback that can change the morphology of a gaseous wake significantly, even altering the sign of the net force \cite[e.g.][and references therein]{2020MNRAS.492.2755G}.
 
%%%%%%%%%%%%%%%%%%%%%%%%%%%%%%%%%%%%%%%%%%%%%%%%%%

\subsection{Numerical Integration}
We solve Eq.~\eqref{EoM} with a simple leapfrog algorithm applying either the hydrodynamic drag or the dynamical friction as the source of the external force to reconstruct the orbital path. The $\vec{r}_{i}$ position of the object is stepped by the following algorithm:
\begin{align}
    \vec{v}_{i+1/2} &= \vec{v}_i + \vec{a}_i \dfrac{\Delta t}{2} \\
    \vec{r}_{i+1} &= \vec{r}_i + \vec{v}_{i+1/2} \Delta t \\
    \vec{a}_{i+1} &= -\dfrac{G m_\star}{\vec{r}_{i+1}^2} + \dfrac{\vec{F}_\mathrm{ext}}{m}\\
    \vec{v}_{i+1} &= \vec{v}_{i+1/2} + \vec{a}_{i+1} \dfrac{\Delta t}{2} 
\end{align}
where $\vec{a}$ is the acceleration and $\Delta t$ is a fixed time-step of the integration. $m_\star = 4\pi \int_0^r \rho(r) r^2 dr$ is the enclosed mass of the envelope at the $r$ position. We integrate the trajectory until the secondary object either (i) approaches a certain fraction of the initial orbital separation e.g.~$10\%$ of the initial semi-major axis, or (ii) the enters the subsonic regime of the dynamical friction. In the dynamical friction simulations, we focus on the eccentricity evolution in the supersonic regime only. After integrating the entire orbital trajectory (up to the above separation criteria), we measure the eccentricity per orbit as $e_i = (r_{a,i} - r_{p,i})/(r_{a,i} + r_{a,i} )$ where $r_{a,i} = \max(r)_i$ and $r_{p,i} = \min(r)_i$ are the local $i$th maxima and minima of the oscillating separation. Similarly, the local semi-major axis is measured as $a_i = (r_{a,i} + r_{p,i})/2$. 

%%%%%%%%%%%%%%%%%%%%%%%%%%%%%%%%%%%%%%%%%%%%%%%%%%

\section{Semi-Analytic Model}
\label{sec:3}

The orbital eccentricity changes due to the external forces that the gas exerts on the secondary as it travels through the medium. The gas--object interaction both leads to the dissipation of the specific orbital energy ($\varepsilon$) and the change in the specific angular momentum $(h)$. We estimate the eccentricity following changes in these values to be
\begin{equation}\label{eq:ecc}
    e \approx e_0 + \Delta e  =  \sqrt{1 + \frac{2(\varepsilon_0 + \Delta \varepsilon) (h_0 + \Delta h)^2}{(GM)^2}} ,
\end{equation}
where $\varepsilon_0 = -\dfrac{GM}{2a_0}$, $h_0 = \sqrt{GMa_0(1-e_0^2)}$ are the initial values of $\varepsilon$ and $h$ with $a$ being the semi-major axis and $e$ being the eccentricity. $M = m_\star + m$ is the total mass. 
 To simplify Eq.~\eqref{eq:ecc}, we introduce $u_0 = e_0^2 -1$ and $u = e^2 - 1$ variables and express the change as 
\begin{equation}\label{eq:du}
    \dfrac{\Delta u}{u_0} \approx \dfrac{\Delta \varepsilon}{\varepsilon_0} + \dfrac{2 \Delta h}{h_0}
\end{equation}
where we neglect quadratic terms in the energy and angular momentum changes as
\begin{equation}
    \frac{(\varepsilon_0 + \Delta \varepsilon) (h_0 + \Delta h)^2}{(GM)^2} \approx \frac{\varepsilon_0 h_0^2}{(GM)^2}\left[ 1 +  \dfrac{\Delta \varepsilon}{\varepsilon_0} + \dfrac{2 \Delta h}{h_0}
\right] \,.
\end{equation}

The source of the energy change is the dissipative external force
\begin{equation}\label{epsilon}
    \Delta\varepsilon = \frac{\vec{v}\cdot \vec{F}_\mathrm{ext}}{m} \Delta t,
\end{equation}
where $\Delta t$ is the time window in which we estimate the transfer of orbital energy to the gas. Similarly, the loss of specific angular momentum comes from the torque that the gas exerts on the companion for $\Delta t$ time
\begin{equation}\label{ha}
    \Delta h = \frac{|\vec{r}\times \vec{F}_\mathrm{ext}|}{m} \Delta t.
\end{equation}

For the approximation of Eq.~\eqref{eq:ecc} to hold, we need to choose a time interval, $\Delta t$, such that $|\Delta \varepsilon/\varepsilon_0|\ll 1$ and $\Delta h/h_0 \ll 1$. In that case the resulting change in eccentricity will be small and the expansion of Eq~\eqref{eq:ecc} will be justified. In practice, we use the orbit as a representative time unit. For systems with slowly-varying orbital elements due to dynamical friction (equivalently where $|F_{\rm ext}/F_{\rm grav}|\ll 1$) changes over an orbital period will be small enough to apply  Eq~\eqref{eq:ecc}.

Additionally, for an eccentric orbit, the gaseous conditions and relative motion of the object through the gas change as a function of orbital phase. Therefore, applying the local conditions at a single position cannot represent the whole orbit \citep[e.g.][]{Salcedo2019}. Instead, we consider the local conditions at the two extremes of periapse and apoapse. Thus, 
\begin{equation}
\Delta\varepsilon \approx \Delta\varepsilon_{\rm p} + \Delta\varepsilon_{\rm a}
\end{equation}
and
\begin{equation}
    \Delta h \approx \Delta h_{\rm p} + \Delta h_{\rm a}
\end{equation} where we introduce the lower index notation for periapsis ($p$) and apoapsis ($a$). To do so, we must approximate the time spent at the orbital locations of periapse, $\Delta t_p$, and apoapse $\Delta t_a$. 

Given these approximations, we calculate $e$ of Eq.~\eqref{eq:ecc} using Eq.~\eqref{epsilon} and Eq.~\eqref{ha} for hydrodynamic drag and gas dynamical friction forces, respectively in the sections that follow. 

%%%%%%%%%%%%%%%%%%%%%%%%%%%%%%%%%%%%%%%%%%%%%%%%%%

\subsection{Hydrodynamic drag}

In case of a pure hydrodynamic drag, the energy dissipation and the change of angular momentum are
\begin{equation}
    \Delta \varepsilon =  - \dfrac{C_d}{2 m}\pi\rho R^2 v^3 \Delta t \; ,
\end{equation}
and
\begin{equation}
    \Delta h = -\dfrac{C_d}{2 m}\pi\rho R^2 |\vec{r}\times\vec{v}^2| \Delta t \; .
\end{equation}
We evaluate these terms only at the peri- and apoapsis where the position is \begin{equation}
    r_\mathrm{p,a} = a(1\mp e)
\end{equation}
and the velocity is 
\begin{equation}
    v_\mathrm{p,a} = \sqrt{\dfrac{GM(1\pm e)}{a(1 \mp e)}}.
\end{equation}
 At these locations in the orbit, the position and velocity are perpendicular. We substitute these expressions into Eq.~\eqref{eq:du}:
\begin{equation}
    \dfrac{\Delta u_{\rm p,a}}{u_0} = \alpha \rho a_0 \left( \dfrac{1\pm e_0}{1\mp e_0} - 1 \right)
\end{equation}
where we introduced $\alpha = \pi R^2 \dfrac{C_d}{m} $ and applied the approximation that $\Delta t_{\rm p,a} \approx a/v_{\rm p,a}$. The eccentricity at the peri- and apoapsis is
\begin{equation}
    e_{\rm p,a} = \sqrt{\alpha \rho a_0 [ 1-e_0^2 - (1 \pm e_0)^2  ] -e_0^2 }.
\end{equation}
 This yields a net eccentricity following a complete orbit of $e \approx e_{\rm p} + e_{\rm a} $.

%%%%%%%%%%%%%%%%%%%%%%%%%%%%%%%%%%%%%%%%%%%%%%%%%%

\subsection{Dynamical friction}

Similarly, we outline the effect of gaseous dynamical friction on the eccentricity evolution. The specific orbital energy and the specific angular momentum changes are
\begin{equation} \label{eq:Deps}
    \Delta \varepsilon =  -4\pi G^2 m\rho I \frac{1}{v}  \Delta t
\end{equation}
\begin{equation} \label{eq:Dh}
    \Delta h = -4\pi G^2 m \rho I \frac{|\vec{r}\times\vec{v}|}{v^3} \Delta t
\end{equation}
We evaluate these terms at the apses again, substitute them to Eq.~\eqref{eq:du}:
\begin{equation}
    \dfrac{\Delta u_{\rm p,a}}{u_0} = \beta \rho a_0^3 \xi \left[ \dfrac{1\mp e_0}{1 \pm e_0} -  \left(\dfrac{1\mp e_0}{1 \pm e_0}\right)^2 \right]
\end{equation}
where we introduced $\beta = 8 \pi m I / M^2$ and $\Delta t_{\rm p,a} = \xi a/v_{\rm p,a}$. We note that $\xi \approx 1$ but can be fit by comparing the prediction with the numerical results. Then
\begin{equation}
    e_{\rm p,a} = \sqrt{\beta \rho a_0^3 \xi \left[ \dfrac{(1\mp e_0)^3}{1 \pm e_0} - (1\mp e_0)^2\right] - e_0^2}
\end{equation}
is the  eccentricity at the peri- and apoapsis. Just as in the case of hydrodynamic drag, the net eccentricity after an orbit can be computed as $e \approx e_{\rm p} + e_{\rm a} $. 

%%%%%%%%%%%%%%%%%%%%%%%%%%%%%%%%%%%%%%%%%%%%%%%%%%

\section{Results}
\subsection{Isothermal, Power-law Density Distribution}
\label{sec:isothermal}

We use the numerical simulations and the above semi-analytic formalism to measure and predict the rate of eccentricity change of a secondary object in an isothermal gas sphere of mass $m_\star$ as a model of an extended primary object. The density profile of the gas is chosen to be a power-law function of the radius as $\rho(r) = \rho_0 r^{-\gamma}$ where
\begin{equation}
    \rho_0 = \frac{(3-\gamma) m_\star}{4 \pi \left(R_\star^{3-\gamma} - R_c^{3-\gamma}\right)} \; ,
\end{equation}
$R_c$ and $R_\star$ are the inner and outer edges of the density profile. We use $R_\star = 1 \, \mathrm{au}$ as the unit of distance and the total enclosed mass at $R_\star$ as the unit of mass $m_\star(R_\star) = 1 M_\odot$ and also $R_c = 10^{-5} R_\star$.

\begin{figure}
\centering
\includegraphics[width=1.0\columnwidth]{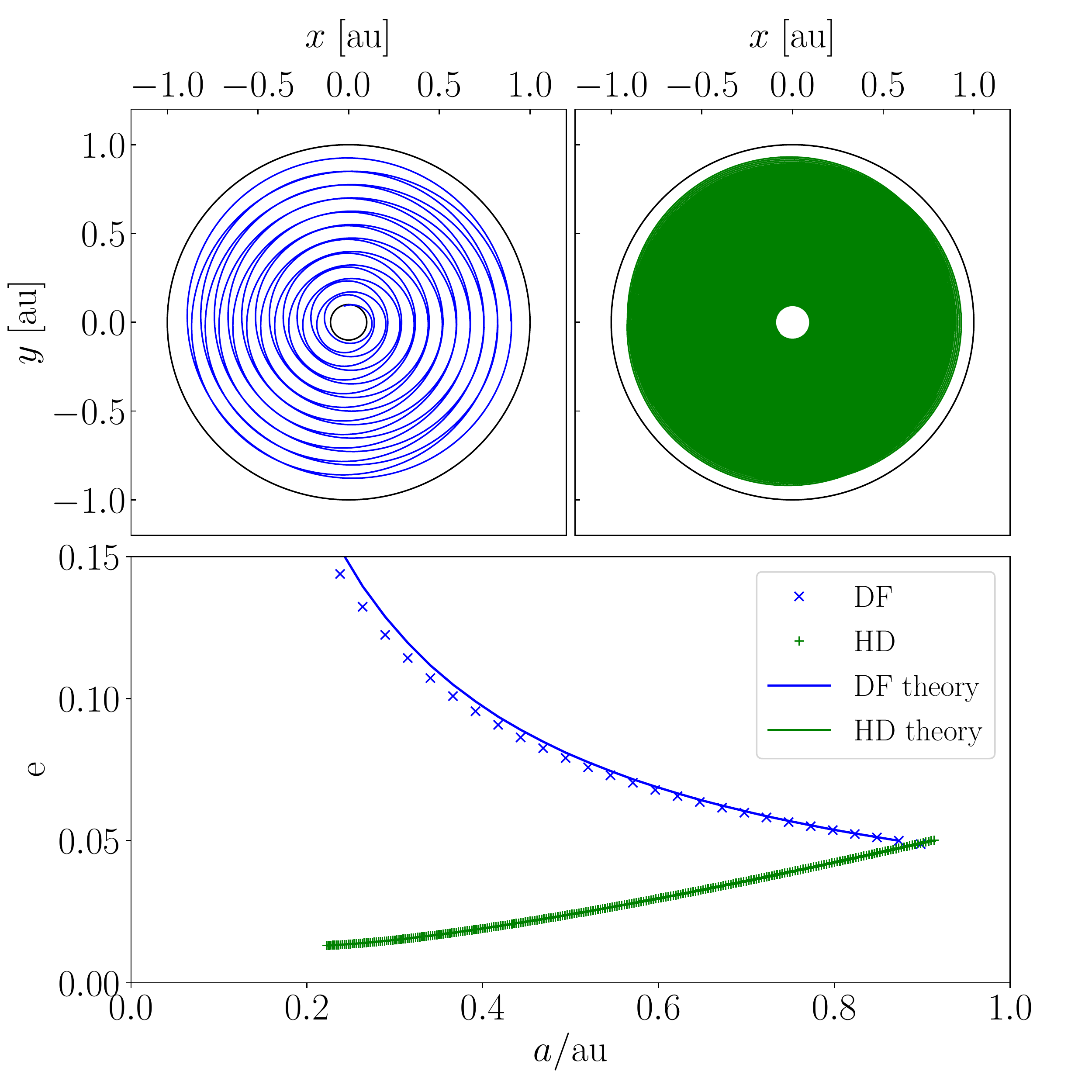} \vspace{-15pt}
\caption{\label{fig1_isotermal_175} \textit{Top:} Planar view of the orbital path of an orbiter in a isothermal gas sphere with a power-law density profile with $\rho(r) = \rho_0 r^{-1.75}$. The companion spirals inward due to gas dynamical friction (DF with blue) on the left panel, and due to hydrodynamic drag force (HD with green) on the right panel, respectively, until the orbit reaches $0.1 a_0$ in both cases. \textit{Bottom:} Orbital eccentricity as a function of semi-major axis calculated per orbit. The mass ratio is $0.001$, the initial eccentricity is $0.05$, and the initial semi-major axis is $0.94 R_\star$ in both DF and HD simulations. Under the influence of dynamical friction, the orbit becomes more eccentric, while under the influence of the hydrodynamic drag it circularizes. }
\end{figure}
In Figure \ref{fig1_isotermal_175}, we show the orbital paths of a  simulation with a $\gamma = 1.75$ isothermal density profile on the top panels. Here, we chose a secondary with mass $m=10^{-3} M_\odot$ with initial semi-major axis $a = 0.94 \, \mathrm{au}$ and initial eccentricity $e=0.05$. The radius of the secondary object is chosen to be $R = 2\cdot10^{-3} \, \mathrm{au}$ and sound speed is a constant as $c_s = \sqrt{1.4 k_B T / m_\mathrm{H2}}$ $\approx 7.6 $ km/s with $T = 10^4 K$, $k_B$ being the Boltzmann-constant, and $m_\mathrm{H2}$ being the mass of molecular hydrogen. The Mach number decreases from its initial value of $\mathcal{M}=3.97$ to $\mathcal{M}=3.31$ (where the simulation was stopped) because the orbital velocity decreases during the inspiral since the enclosed mass is getting smaller as the orbit tightens. These initial parameters set the secondary to orbit within the density profile ($r < R_\star$). The top left and right panels show the simulated inspiral of the secondary object under either purely dynamical friction (with blue) or purely hydrodynamic drag (with green), respectively, in the x-y plane of the motion. Solid circles indicate $R_\star$ (the outer cut-off of the gas) and $0.1 a_0$ (i.e.~$10\%$ of the initial semi-major axis) radii. We ran the simulations between $r = 0.09 - 0.94 \, \mathrm{au}$ radius range. The inspiral times are $t_\mathrm{df} = 11.7$ yr and $t_\mathrm{hyd} = 109.1$ yr, respectively. The bottom panel shows the eccentricity change during the inspiral. With $\gamma = 1.75$ density profile, the dynamical friction causes the orbit to become more eccentric while hydrodynamic drag circularizes the orbit.

\begin{figure}
\centering
\includegraphics[width=1.0\columnwidth]{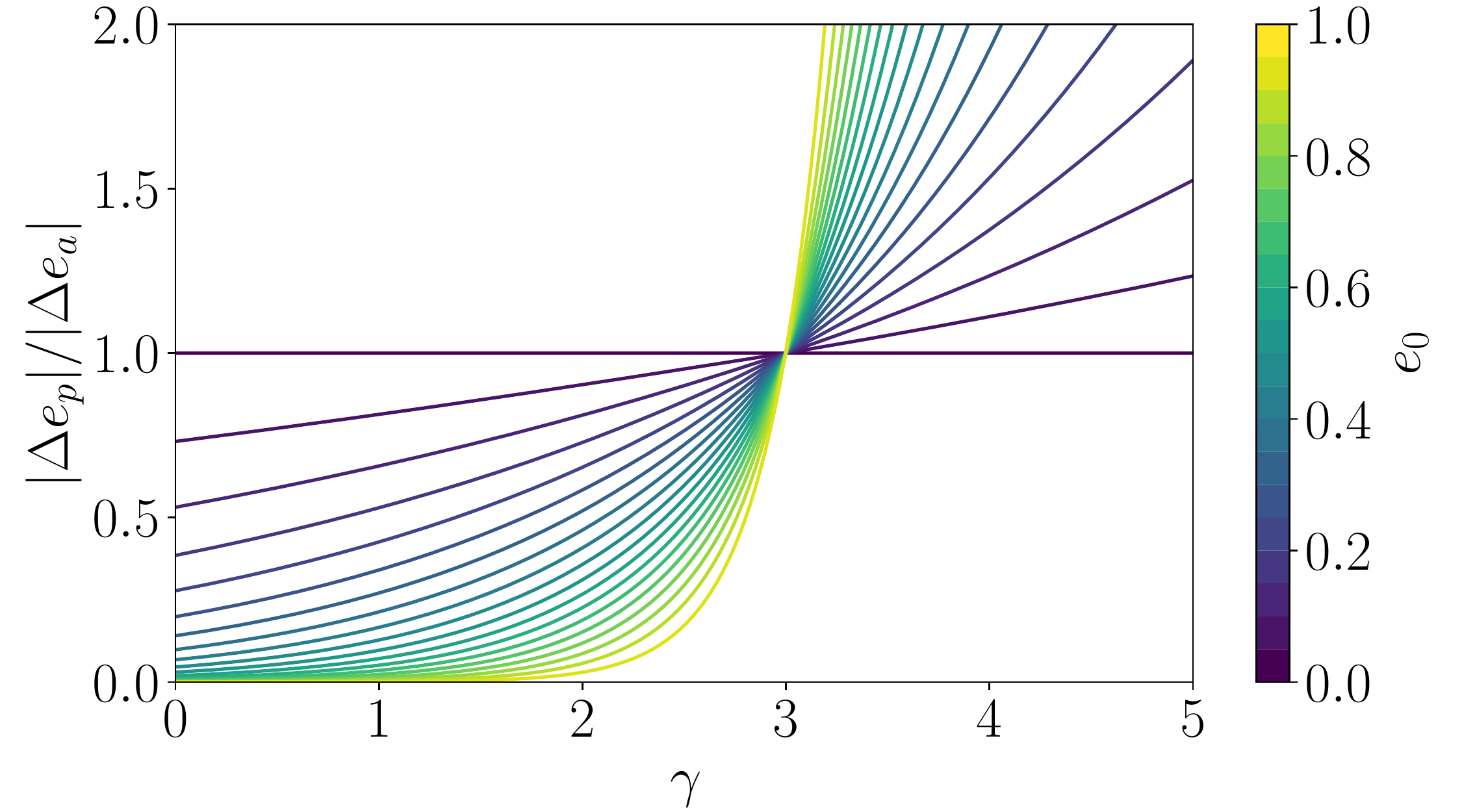} \vspace{-15pt}
\caption{\label{gamma_isothermal} The ratio between the eccentricity change at peri- and apoapses as a function of the slope ($\gamma$) of the isothermal density profile in dynamical friction. The ratio of change of the orbital eccentricity is shown for 20 different realizations of a system. Colors indicate the initial eccentricities of the 20 systems. For all eccentricities, $\gamma=3$ is a critical value that divides greater eccentricity change at periapse and apoapse. The signs of $\Delta e_p$ and $\Delta e_a$ imply that when $\gamma<3$ orbits become more eccentric, while when $\gamma>3$ they the become less.}
\end{figure}

Using the same mass ratio but $a_0 = 0.5 \, \mathrm{au}$, we further analysed the effect of dynamical friction on the eccentricity evolution for the entire range of initial eccentricities. We predict if the eccentricity is about to be excited or damped as a function of $\gamma$ and $e_0$ at the first orbit for both apses. We show the ratio between $\Delta e_{\rm p} = e_{\rm p} - e_0$ and $\Delta e_{\rm a} = e_{\rm a} - e_0$ in Figure \ref{gamma_isothermal} where the colors indicate the initial eccentricities of the systems from $e_0 = 0$ up to $1$. 

Because forces at periapse tend to circularize the orbit while those at apoapse tend to make it more eccentric, the ratio $\Delta e_{\rm p} / \Delta e_{\rm a}$ is an important indicator of the orbital evolution. We find that $\Delta e_{\rm p} / \Delta e_{\rm a} < 1$ for $\gamma < 3$ which means that the orbits tend to become more eccentric in shallow gas density profiles. By contrast, orbits circularize in cuspy density profiles for $\gamma > 3$. More eccentric the initial orbits are, the more eccentric they get ($\gamma<3$) or the more they circularize ($\gamma>3$). 

\begin{figure}
\centering
\includegraphics[width=1.0\columnwidth]{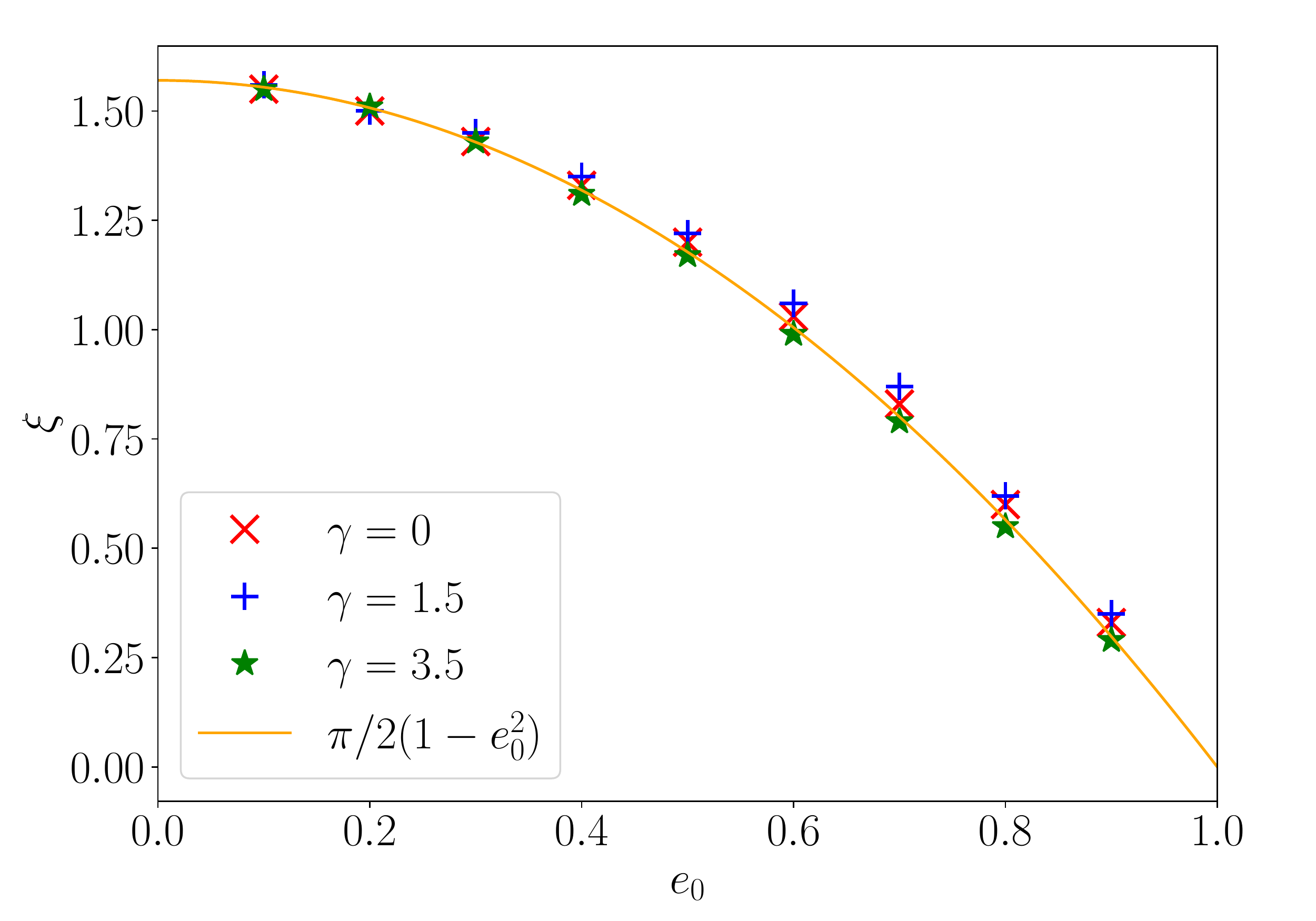} \vspace{-15pt}
\caption{\label{fitting} The fitting parameter $\xi$ on which we approximate the apses transit ($\Delta t = \xi a / v$) as a function of initial eccentricity for $3$ models of power-law density profiles ($\gamma = 0.0,1.5,3.5$) for dynamical friction simulations. }
\end{figure}

We have compared the above semi-analytic results with the numerical simulations and found that $\xi$ scaling parameter can be fit as $\xi \approx {\frac{\pi}{2}} (1-e_0^2)$ in case of dynamical friction. Note that $\xi$ is a fitting parameter that defines the time window $\Delta t = \xi a/ v$ in which we approximate the apses transit in Eq.~\eqref{epsilon} and Eq.~\eqref{ha}. $\xi$ is measured by comparing the discrepancy between the eccentricity change in the numerical simulations versus the eccentricity change in the semi-analytic model with $\xi = 1$. In Figure \ref{fitting}, we show how we measured the deviations at 9 points of initial eccentricities for 3 models ($\gamma = 0,1.5,3.5$).

%%%%%%%%%%%%%%%%%%%%%%%%%%%%%%%%%%%%%%%%%%%%%%%%%%

\subsection{Common Envelope Inspiral}
\label{sec:4}

An ubiquitous feature of simulations of hydrodynamic simulations of common envelope inspiral is the development of moderate eccentricity -- even when these models are initialized in circular orbits. This is seen clearly, for example in Figure 2 of \citet{2012ApJ...746...74R}, Figure 11 of \citet{2012ApJ...744...52P} and Figures 1 of \citet{2016ApJ...816L...9O} and \citet{2019MNRAS.490.3727C}. The analysis of \citet{2019MNRAS.490.3727C} progresses much further and analyses the drag forces applied to the infalling bodies, and discusses the similarity (in the early inspiral) and departure (after the envelope has been disturbed) of these forces from those predicted by gaseous dynamical friction and an undisturbed profile. Yet, results appear quite varied across the models presented by different groups and initial conditions. For example, the eccentricity seen in the models of \citep{2020A&A...644A..60S} is quite small compared to that of \citet{2016ApJ...816L...9O} when the same code but different initial conditions are employed. To our knowledge, no clear explanation for this eccentricity, or why it differs across simulation models, has been outlined. We explore this question further in this section by examining the eccentricity evolution in a subset of hypothetical model common envelope inspiral episodes. 

We simulate the eccentricity evolution of a Jupiter-like planet engulfed by a Sun-like giant star at its late, red giant evolutionary phase \citep[e.g.][]{1984MNRAS.210..189S,1998ApJ...506L..65S,1999MNRAS.304..925S,1999MNRAS.308.1133S,2009ApJ...700..832C,2012MNRAS.425.2778M,2014ApJ...787..131Z,2016MNRAS.458..832S,2016ApJ...829..127A,2018ApJ...853L...1M,2018AJ....156..128S,2018ApJ...864...65Q,2019MNRAS.490.2390P,2020ApJ...889...45S,2020JCAP...10..027J,2021AJ....162..273S,2021MNRAS.507.2659G,2022arXiv220311227Y}. The total mass of the star is $m_\star = 1 M_\odot$ and its radius is $R_\star = 1 \, \mathrm{au}$, the planet has a mass of $m = 10^{-3} M_\odot$ and a radius of $R = 2\cdot10^{-3} \, \mathrm{au}$ (or approximately 4.28 Jupiter radii). The star comprises two structural parts: the core which makes up $1/3$ of its total mass and a polytropic envelope which extends up to $R_\star$ and has a polytropic index $n = 1.5$. For the sake of simplicity, we treat the stellar core as a point mass in the centre, and construct the envelope solution according to the Lane-Emden equation for $n=1.5$ polytropic index:
\begin{equation}\label{eq:LE}
    \dfrac{1}{\zeta^2} \dfrac{d}{d\zeta} \left( \zeta^2 \dfrac{d\Theta}{d \zeta} \right) = -\Theta^n \, ,
\end{equation} 
where $\zeta$ is a distance parameter. We numerically integrate this differential equation and express the density as:
\begin{equation} \label{eq:dens}
    \rho(r) = \frac{\max(\zeta)^3}{8/3 \pi \int_0^{\max(\zeta)} \Theta^n(\zeta) \zeta^2 d\zeta}  \Theta^n(r) \, ,
\end{equation} where we apply the variable transformation $r = \zeta / \max(\zeta)$ such that $\max(\zeta)$ is the zero of the $\Theta(\zeta)$ function. This normalization meets the requirement that the total mass of the envelope is $2/3$ and its radius is $1$  since we chose $G=M_\mathrm{unit}=R_\mathrm{unit}=1$ unit system, which can be rescaled to any physical dimension. We note that the inner boundary condition of a point-mass core does not affect the orbital evolution of the planet through its inspiral in the stellar envelope because the planet's acceleration only depends on the gravity of the enclosed mass and the local density of the envelope in our simplified model. 
 
In case of dynamical friction however, the orbital evolution also strongly depends on the system's Mach-number. We only examine the first, rapid evolutionary phase of the orbit where the planet orbits in the supersonic regime i.e.~$v > c_s$. Here, we evaluate the corresponding $I$ parameter of Eq.~\eqref{df} locally as $I_\mathrm{sup} = 0.5\ln ( 1 -[v/c_s]^{-2}) - \ln \left( r / R\right)$. The second term is also know as the Coulomb-logarithm in which $r$ is the separation between the planet and the stellar core. In the first term, we approximate the local sound speed as $c_s = c_0 \sqrt{(1+1/n) \rho^{1/n}(r)}$ where $c_0 = 29.7 km/s$ is a unit conversion parameter.  

\begin{figure}
\centering
\includegraphics[width=1.0\columnwidth]{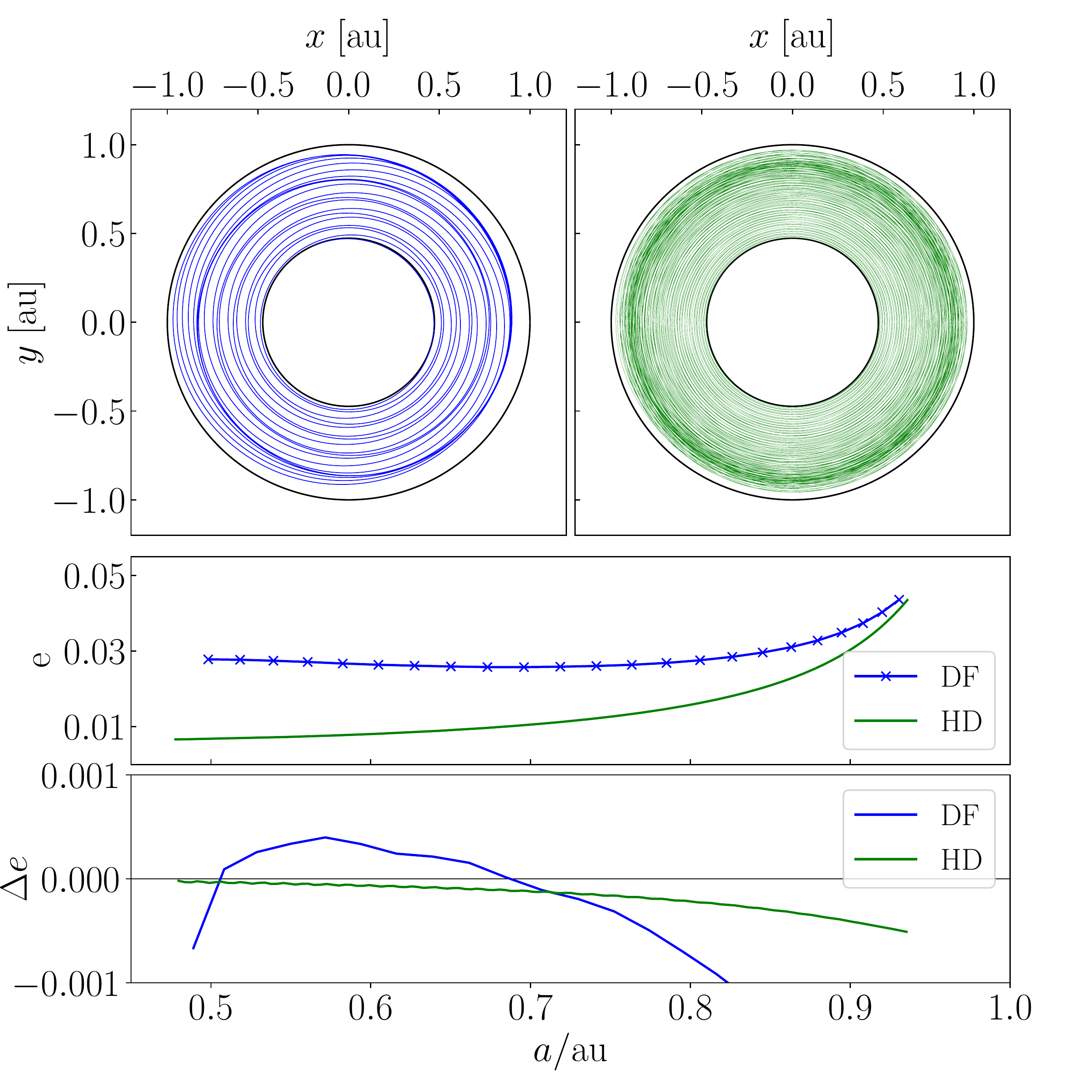} \vspace{-15pt}
\caption{\label{app1} \textit{Top:} Planar view of the orbital path of a companion in a polytropic envelope of a giant star with $n = 1.5$ polytropic index and $R_\star$ radius. The companion spirals inward due to gas dynamical friction (DF with blue) on the left panel, and due to hydrodynamic drag force (HD with green) on the right panel, respectively, until the orbit reaches the subsonic regime of dynamical friction i.e. where $v(r) = c_s(r)$ (for which $r \approx 0.5 R_\star$). \textit{Bottom:} Orbital eccentricity as a function of semi-major axis calculated per orbit. The mass ratio is $0.001$, the initial eccentricity is $0.05$, and the initial semi-major axis is $0.94 R_\star$ in both DF and HD simulations.}
\end{figure}
In Figure \ref{app1}, we show the orbital paths of the planet on the top panels (similarly to Figure \ref{fig1_isotermal_175}). The left panel shows the case in which the planet's orbit evolves under the effect of dynamical friction only. Similarly, the right panel shows the same hypothetical case where the orbit of the planet shrinks under the influence of hydrodynamic drag only. The simulations were started at $a_0 = 0.94 \, \mathrm{au}$ with $e_0 = 0.05$ initial eccentricity. In both cases, the planet migrates inward and its orbit start to circularize. The inspiral times are $t_\mathrm{df} = 13.1$ yr and $t_\mathrm{hyd} = 96.1$ yr, respectively. In case of dynamical friction, we observe a transitional phase in which the circularization stops between $a \approx 0.5 - 0.7 \, \mathrm{au}$ where the planet's orbit becomes more eccentric. These changes reflect the changing steepness of the density profile within the polytropic structure -- near the stellar limb, the scale height is small and the density gradient is very steep, leading to orbital circularization. Deeper in the stellar interior, the scale height can be much larger and the orbit becomes more eccentric under the influence of the shallow density profile. Just before entering the subsonic regime (at which point we stop the simulation), the dynamical friction force drops as the Mach-number approaches unity $F_\mathrm{df} \propto \ln(1-\mathcal{M}^{-2})$. At this limit, the orbit starts to circularize again that continues in the subsonic regime. 

\begin{figure}
\centering
\includegraphics[width=1.0\columnwidth]{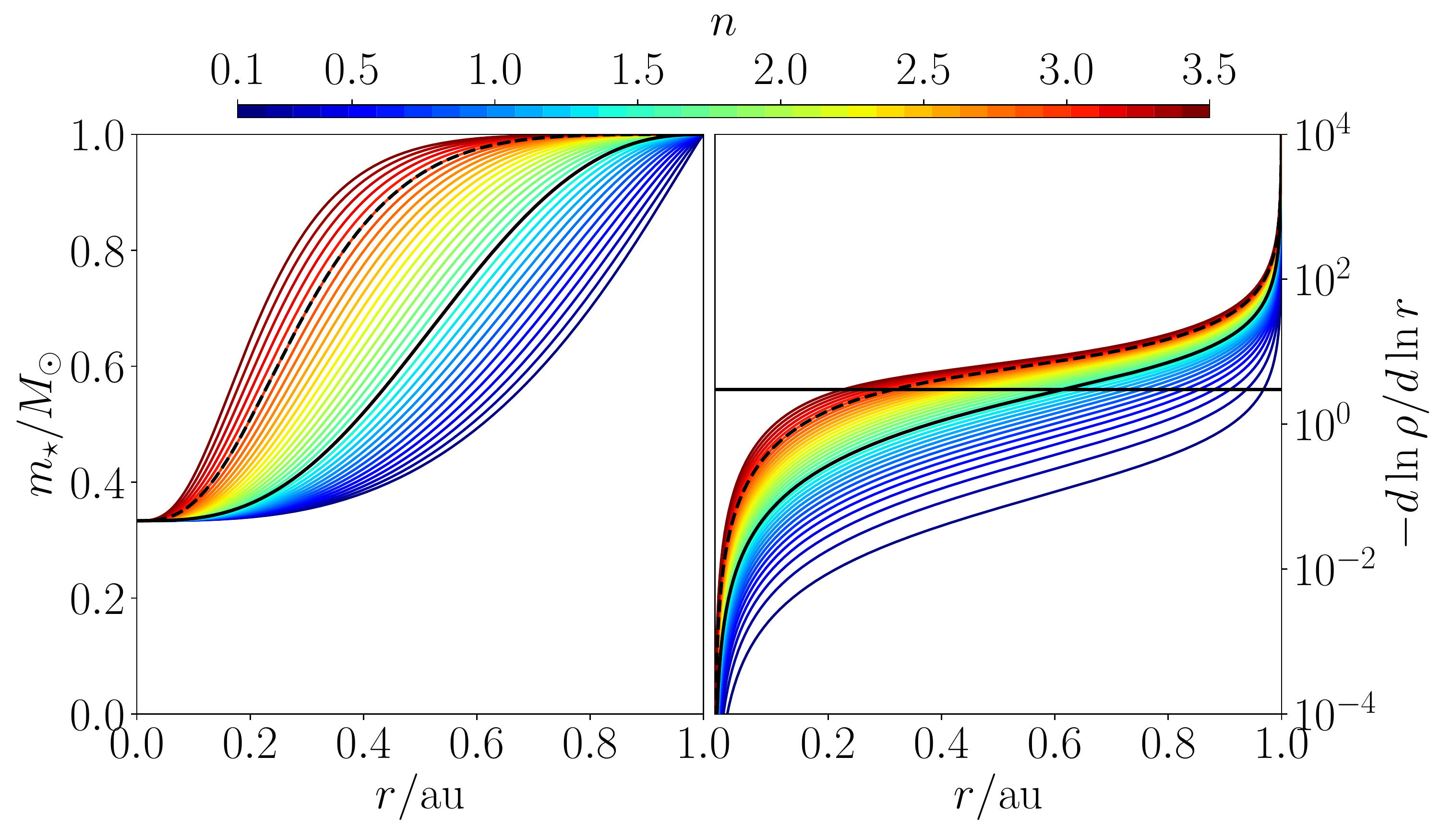} \vspace{-15pt}
\caption{\label{encmass}\textit{Left:} The enclosed mass of the stellar envelope as a function or radius for different polytropic model. \textit{Right:} The $d \ln \rho / d \ln r$ derivative of the density as a function of radius. A horizontal line compares to the critical value of $-d \ln \rho / d \ln r=3$ identified in the context of power-law density profiles. The simulated range of polytropic indexes varies from $n = 0.1$ up to $n = 3.5$ see the colorbar. Important $n = 1.5$ and $3$ models are emphasized with solid and dashed black curves. Note that simulations adopted a point mass-like stellar core with mass $1/3$ of the total mass.}
\end{figure} 
After investigating one specific configuration, we extend our analysis to understand how the observed transition point in the eccentricity evolution depends on the density profile of the stellar envelope. We generate $35$ independent density profiles with polytropic indexes ranging from $n = 0.1$ up to $n = 3.5$ evenly. We show the enclosed masses of the different polytropic envelopes with colored curves as a function of radius on the left panel of Figure \ref{encmass}. We emphasize important polytropes i.e.~$n=1.5$ and $3$ with solid and dashed black curves, respectively. (In all figures from Figure \ref{encmass}, colors and solid and dashed black curves represent the same polytropic indexes of the corresponding stellar envelopes.) In all models, we use the same point mass-like stellar core with mass $1/3 M_\odot$. We also show the slope i.e.~$-d\ln \rho /d \ln r$ of the radial density profiles of each polytropic model as a function of radius on the left panel of Figure \ref{encmass}. The corresponding power-law exponents increase from the core to the surface for polytropes as a function of radius. The diverging exponents at the boundaries are the consequences of the smooth transition feature of the polytropic density profiles at the core and the surface. For comparison, the horizontal black line shows $-d\ln \rho /d \ln r = 3$ which corresponds to the critical $\gamma = 3$ case of the power-law density profiles which separates the quality of eccentricity evolution in isothermal models. 

\begin{figure}
\centering
\includegraphics[width=1.0\columnwidth]{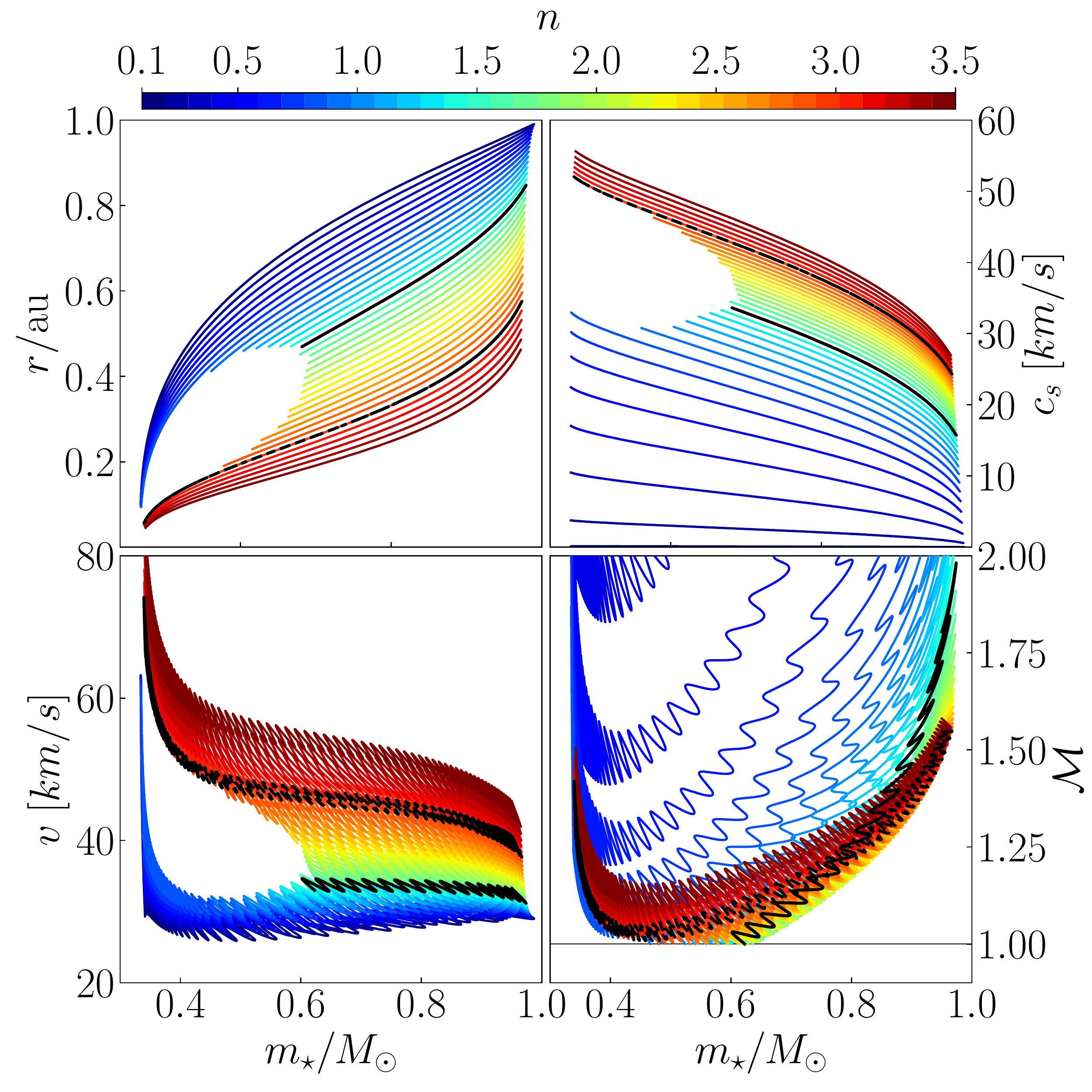} \vspace{-15pt}
\caption{\label{rvcm} \textit{Left:} The orbital separation (top) and velocity (bottom) of the planet as a function of enclosed mass for a series different of polytropic models indicated by the colormap. \textit{Right:} The local sound speed (top) and Mach number (bottom) as a function of enclosed mass for the same series of polytropic models. Solid and dashed black curves emphasize models with $n=1.5$ and $3$. Simulations are stopped either when the planet reach $10\%$ of the initial semi-major axis or enter the subsonic regime in the envelope.}\vspace{-5pt}
\end{figure}
We run $35$ simulations that are shown in Figure \ref{rvcm} using the same initial eccentricity ($e_0 = 0.05$) for the planet but different stellar envelope profiles and different initial semi-major axes. We set the initial semi-major axes in a way that the initial periapsis is at the radius corresponding to $0.95 M_\odot$ enclosed mass. Panels on the right show the separation (top) and the velocity (bottom) of the planet as a function of enclosed mass. Simulations are stopped either when the planet reaches $10\%$ of the initial semi-major axis (e.g.~$ n \lesssim 0.5$ or $3.2 \lesssim n$) or enter the subsonic regime (e.g.~$0.5 \lesssim n \lesssim 3.2$) in the envelope.

\begin{figure}
\centering
\includegraphics[width=1.0\columnwidth]{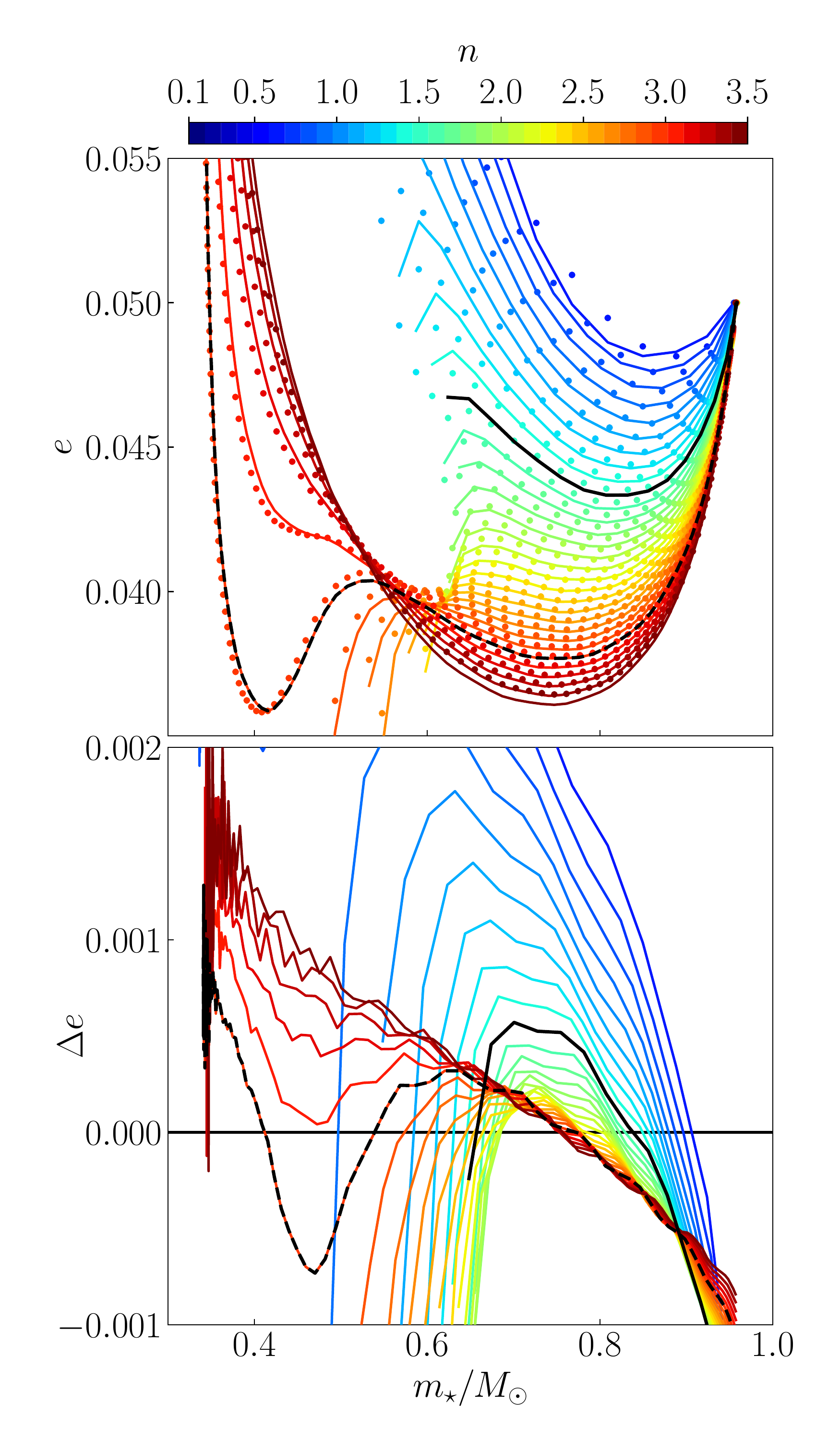} \vspace{-15pt}
\caption{\label{eccentricity}  \textit{Top:} Eccentricity as a function of enclosed masses during the inspiral of the planet in the stellar envelope. Color scale labels the polytropic index of the $35$ different models. Solid and dashed black curves emphasize important $n=1.5$ and $3$ models. Dots show the measured eccentricities per orbit from the numerical simulations while solid curves show the corresponding semi-analytic prediction. \textit{Bottom:} The differential values of the data from numerical simulation from the top panel i.e.~the eccentricity change between two consecutive orbits in the given polytrope. The horizontal black line separates regimes in which the orbit circularize (below zero) or becomes more eccentric (above zero).}
\end{figure} 
Figure \ref{eccentricity} shows our results regarding the eccentricity evolution of a planet's orbit during its inspiral in various models of a stellar envelope. The top panel shows the eccentricity as a function of enclosed mass in the numerical simulation (dots) and in the semi-analytic estimate (solid curves). Here, we calculate $e_{i+1}$ by taking the local $(\varepsilon_0,h_0)$ values of the numerical simulation in Eq.~\eqref{eq:ecc}. Similarly we calculate $\Delta \varepsilon$ of Eq.~\eqref{eq:Deps} and $\Delta h$ of Eq.~\eqref{eq:Dh} under the local conditions in each model. The bottom panel shows the differential values from the top panel i.e.~$\Delta e_i = e_{i+1} - e_i$ of the numerical simulation. The horizontal line indicates $\Delta e = 0$ below which orbits tend to circularizes and above which orbits become more eccentric. Our results show that after initial circularization the orbital eccentricity start to increase in various models before either approaching $0.1 a_0$ or entering the subsonic regime. This is most prominent for polytropes with $n \lesssim 2$ or $n \gtrsim 3$ which simulations never enters the subsonic regime but appears in all models. 

\begin{figure}
\centering
\includegraphics[width=1.0\columnwidth]{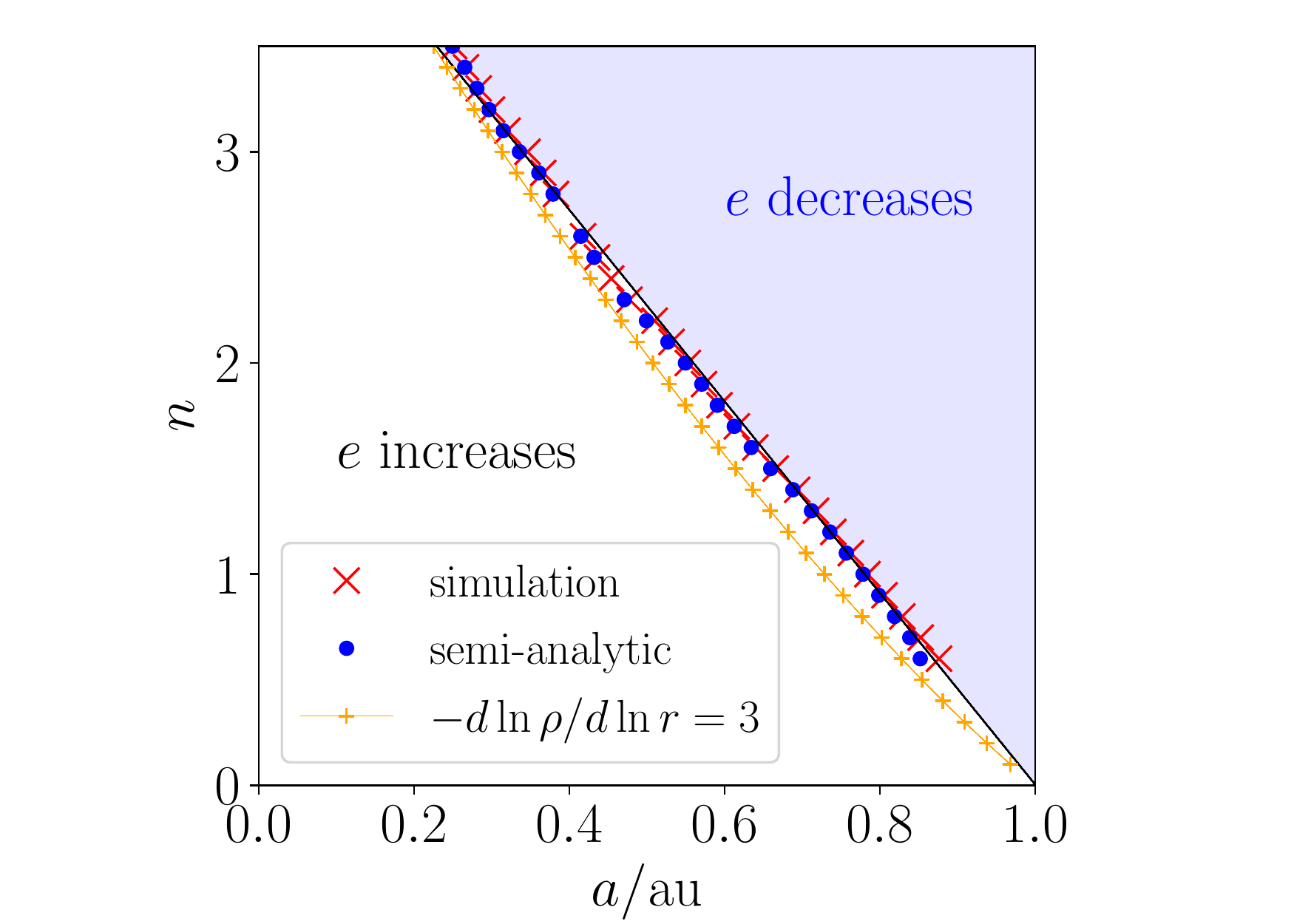} \vspace{-15pt}
\caption{\label{app3}The transition points from circularization to eccentricity growth in different polytropes as function of semi-major axis as a fraction of the stellar radius. Both numerical results (with red $\times$-s) and semi-analytic predictions (with blue dots) are shown together with a linear fit (black solid line) to the numerical results. The orange curve shows the transition points corresponding to $d \ln \rho / d \ln r = 3$ in Figure \ref{encmass}.}
\end{figure}

In our model polytropes, we measure the first transition points where the initial circularization turns into eccentricity growth as a function of semi-major axis, Figure \ref{app3}. We observe that our numerical integrations and semi-analytic predictions are in very close agreement, and trace a relationship that defines the fractional critical radius as a function of the envelope's polytropic index. We note, however, that polytropes with a different core mass fraction would lead to a different result. We note that our result can be fit with the linear relation $n(a/R_\star) = -4.53 a/R_\star + 4.54$.  Finally, we compare this relationship to a criterion based on the local density slope, $d\ln \rho / d\ln r = -3$. We find that this approximation, while not as accurate as the full semi-analytic theory (because it neglects the dependence of the dynamical friction drag force on the local Mach number), provides useful context for the resulting eccentricity evolution. 

Given these findings, we argue that the ubiquitous presence of eccentricity in simulated common envelope inspirals can be traced to gaseous dynamical friction on the extended envelope. In the early inspiral, the steep gradient of the outer envelope ($-d\ln \rho / d\ln r  \gtrsim3$) damps any orbital eccentricity,  while in the later inspiral, eccentricity can be enhanced by interaction with the comparatively homogeneous envelope interior ($-d\ln \rho / d\ln r \lesssim 3$).  

%%%%%%%%%%%%%%%%%%%%%%%%%%%%%%%%%%%%%%%%%%%%%%%%%%

\section{Conclusions}
\label{sec:conclusion}

In this paper, we examined the eccentricity evolution of an object orbiting in an extended gaseous medium due to hydrodynamic drag or gas dynamical friction. We built a simple numerical integrator to calculate the orbital path of the companion under the influence of the gravity of the primary and the frictional forces exerted by the gas. We measured the eccentricity evolution during the inspiral of the companion and compared the results with the prediction of a semi-analytic approach. We compared the relative importance of hydrodynamic drag and gas dynamical friction in the eccentricity evolution. We focused our analysis on dynamical friction dominated regimes. Some key findings of our study are:
\begin{enumerate}
    \item Drag forces applied to the system at periapse tend to make orbits more circular, while those applied at apoapse tend to make orbits more eccentric. 
    \item In all centrally-concentrated mass distributions, hydrodynamic drag causes orbital eccentricity to decrease because the drag increases with increasing velocity (e.g. at periapse in an eccentric orbit), see equation \eqref{hyd} and Figure \ref{fig1_isotermal_175}.
    \item Because the gaseous dynamical friction drag force decreases with increasing velocity in the supersonic regime, equation \eqref{df}, whether orbits become more or less eccentric under the influence of gaseous dynamical friction depends on the density profile.
    \item We find that the critical value for a radial power-law density distribution $\rho \propto r^{-\gamma}$ is $\gamma=3$, where lower values of $\gamma<3$ drive orbital eccentricity increase while higher values of $\gamma>3$ drive orbital eccentricity decrease (Figure \ref{gamma_isothermal}). 
    \item Both the sign and rate of change of orbital eccentricity can be be accurately predicted by our semi-analytic theory of section \ref{sec:3} when coupled with the numerical coefficient $\xi \approx \pi/2 (1-e^2)$, as reported in Figure \ref{fitting}. 
\end{enumerate}
We apply this theory of eccentricity evolution under the influence of gaseous dynamical friction to the orbital evolution of engulfed objects in common envelope phases. Here we consider an example of a Jupiter-like planet interacting with the envelope of a Sun-like star at its late, red giant evolutionary phase. The hydrostatic mass distribution of the stellar envelope has a steep density gradient (lower temperature and smaller scale height) near the surface and a shallower density gradient (higher temperature and larger scale height) in the deep interior. We show that as orbiting objects pass through these mass distributions, the experience circularization in the outer envelope, and eccentricity excitation in the inner envelope. For polytropes of varying index, $n$, we demonstrate the eccentricity evolution and the inflection between eccentricity decrease in the outer envelope and increase in the inner envelope (Figure \ref{eccentricity}). The inflection between eccentricity decrease and growth can be modeled accurately with our semi-analytic model, or approximately by finding the radius within the stellar model where $d \ln \rho / d\ln r = -3$ (Figure \ref{app3}).  

Conceptually, our results provide a framework for understanding the evolution of eccentricity in objects being dragged inward in gaseous distributions. In particular, we demonstrate that the development of orbital eccentricity in global hydrodynamic simulations of common envelope phases is indeed realistic, rather than being an artifact of numerics or initial conditions. Similarly, runaway growth of eccentricity was observed in non-gaseous dynamical friction for unequal mass binaries in stellar background \citep[e.g.][]{2012MNRAS.422..117M}. Our semi-analytic model adopts the Mach-number dependent coefficients of \citet{Ostriker1999}, but could equally be extended to coefficients of dynamical friction that depend on the local density gradient or other properties \citep[e.g.][]{MacLeod2017,2020ApJ...897..130D}. Our results are further suggestive that the emergence of objects from common envelope phases with moderate, non-zero eccentricities may be a natural consequence of the physics of gaseous dynamical friction. There are many other applications of our results, including the formation of Thorne-Zytkow objects \citep[][]{1975ApJ...199L..19T}, the migration of stars in accretion flows around black holes, and the prediction of gravitational wave emissions of eccentric compact binaries \citep[see e.g.][]{2013ApJ...774...48M,2020MNRAS.493.4861G,2021PhRvD.103b3015C}. Our model can be extend towards including feedback which could have potential effect on the orbital evolution by damping or reversing the gaseous dynamical friction \cite[see e.g.][]{2020MNRAS.492.2755G} in certain astrophysical scenarios with high outflow rates.

%%%%%%%%%%%%%%%%%%%%%%%%%%%%%%%%%%%%%%%%%%%%%%%%%%

\vspace{-15pt}
\section*{Acknowledgements}

We thank Vitor Cardoso, Rodrigo Vicente, Ari Laor and Yohai Meiron for helpful comments. The project was sponsored by a Fulbright Student Grant through the Hungarian - American Commission for Educational Exchange. This work was supported by the National Science Foundation under Grant No. 1909203 and by the Black Hole Initiative - which is supported by grants from JTF and GBMF.

%%%%%%%%%%%%%%%%%%%%%%%%%%%%%%%%%%%%%%%%%%%%%%%%%%
\section*{Data Availability}

The python scripts used to create and analyse our simulations and models are available from the corresponding author upon request and publicly available from http://galnuc.elte.hu/Eccentricity\_Evolution\_in\_Gaseous\_Dynamical\newline \_Friction.zip

%%%%%%%%%%%%%%%%%%%% REFERENCES %%%%%%%%%%%%%%%%%%

% The best way to enter references is to use BibTeX:

\vspace{-15pt}
\bibliographystyle{mnras}
\bibliography{szolgyen_macleod_loeb}

%%%%%%%%%%%%%%%%%%%%%%%%%%%%%%%%%%%%%%%%%%%%%%%%%%

% Don't change these lines
\bsp	% typesetting comment
\label{lastpage}
\end{document}